\documentclass[twocolumn,showpacs,preprintnumbers,amsmath,amssymb]{revtex4}

\usepackage{graphicx}% Include figure files
\usepackage{dcolumn}% Align table columns on decimal point
\usepackage{bm}% bold math

\begin{document}

%\preprint{APS/123-QED}

\title{Vlasov model using kinetic phase point trajectories}

\author{H. Abbasi$^{a}$}
 \email{abbasi@aut.ac.ir}
\author{M. H. Jenab$^{a}$}%
\author{H. Hakimi Pajouh$^{b}$}
\affiliation{
$^{a}$Faculty of Physics, Amirkabir University of Technology,\\
P. O. Box 15875-4413, Tehran, Iran\\
$^{b}$Department of Physics, Alzahra University,\\
Tehran 19834, Iran.
}

\begin{abstract}
A method of solution of the collisionless Vlasov equation, by following collisionless phase point trajectories in phase space, is presented. It is shown that by increasing the number of phase points, without enhancing the resolution of phase space grid, the accuracy of simulation will be improved. Besides, the phase points spacing introduces a smaller scale than grid spacing on which fine structures might be more conveniently handled. In order to perform simulation with a large population of phase points, an effective interpolation scheme is introduced that reduces the number of operations. It is shown that by randomizing initial position of the phase points along velocity axis, the recurrence effect does not happen. Finally, the standard problem of linear Landau damping will be examined.      

\end{abstract}

\pacs{83.85.Pt, 52.65.Ff, 82.20.Fd, 52.35.-g}


\maketitle
There are two equally important numerical approaches to the kinetic problem. The first approach, particle-in-cell (PIC) methods, self-consistently models the plasma by a finite number of macroparticles on a fixed grid. Its advantage is replacing the Vlasov equation by the ordinary differential equations of motions of macroparticles that makes PIC codes easily extendable to multi-dimensional applications. However, it is well-known that the numerical noise (proportional to $1/\sqrt{N}$ where $N$ is the number of macroparticles) inherent to PIC simulation becomes, in some cases, too significant to allow a precise description of the distribution function (DF).  Nevertheless, the PIC approach can produce accurate results when a sufficiently large number of macroparticles involve in the simulation \cite{Birdsall}. 

The second approach is direct solution of the Vlasov equation (for a collisionless plasma) that is noise-free in comparison to PIC simulation. The major problem of the Vlasov simulation has been the development of fine structures (filamentation) in velocity space; i.e., a problem with no seemingly simple cure. Partial treatments such as increase in velocity resolution, have sharply limited the ability to extend the above work to higher dimensions \cite{Buchner,Kazeminezhad,Bertrand} and thus treat realistic problems. A large class of the Vlasov simulation models is based on discretizing the Vlasov equation (mainly by a splitting scheme) on a phase space fixed grid. In the splitting method, the new $f$ was obtained as an algebraic expression in terms of the old $f$ by a suitable interpolation method \cite{Johnson}. There are a couple of problems with the splitting scheme: (i) it was not rigorously shown under which circumstances the coupled equations have solutions ''approximately'' consistent with the Vlasov equation; and (ii) following characteristics along the phase space coordinates departs one from the characteristics on which $f$ truly remains invariant \cite{Kazeminezhad}. Besides, for them the partial treatment of the filamentation, i.e. increasing the velocity resolution, is very expensive because all of the simulation operations is performed on the grid. Moreover, they are suffering from the recurrence effect \cite{Shukri}.   

Integration of the Vlasov equation along the collisionless phase point trajectories has been the most promising of these methods \cite{Kazeminezhad}. This method is based on following the characteristics along which $f$ is constant in the collisionless case. Therefore, characteristic equations of the Vlasov equation are solved. That is one of the advantages of the scheme and makes it possible to use all the PIC simulation experiences. A complete description of the method of characteristics was presented in Ref. \cite{Kazeminezhad}. There, besides the grid points there are equal number of phase points over which the DF, $f_p$, is initially defined. Interpolation is performed between the phase points and the fixed background grid to obtain the DF, $f_g$, on the grid. The main advantage of introducing $f_p$ is that it remains unchanged when the phase points follow their characteristics (contrary to the semi-Lagrangian method in which the DF is altered by the interpolation).
 
In present paper, we first improve the accuracy of the simulation by increasing the number of phase points in each grid cell, without enhancing the resolution of the phase space grid. Better accuracy is the result of sampling DF with higher resolution. Larger population of phase points, in comparison to grid points, introduces a smaller scale than grid spacing on which fine structures might be more conveniently handled. However, increasing the number of phase points necessitates an effective interpolation scheme (IS) that reduces the number of operations while keeping the accuracy [{\it e.g.} in comparison to the bilinear interpolation scheme (BIS)]. Accordingly, a new IS is introduced. In such a way that the DF of each grid point is obtained by averaging the DF of phase points located in four cells around a grid point. Moreover, it is shown that by randomizing the initial position of the phase points along velocity axis, the recurrence effect does not happen and the reason is given in detail.      

\begin{figure}
\hskip -4cm
\includegraphics[height=4cm,width=4.6cm]{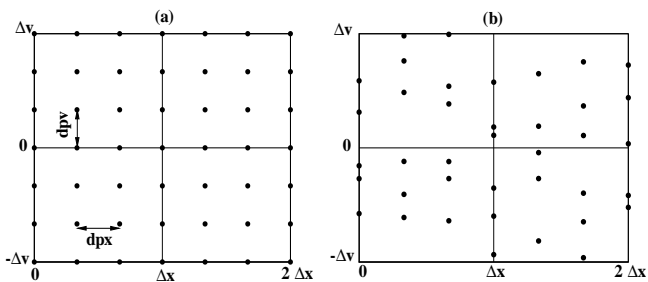} \\
\caption{\label{fig1} A typical part of the phase space grid. (a) Regular arrangement (b) Random arrangement along the velocity direction.}
\end{figure}

Our mathematical model is the one-dimensional Vlasov-Poisson system,
\begin{eqnarray}
&& \partial_t f + v \partial_x f -E(x,t) \partial_v f = 0,  \nonumber\\ 
&& \partial_{x} E = 1 - \int_{-\infty}^{+\infty} f dv, \label{1}
\end{eqnarray}
where $f$ is the electron DF and $E$ is the electric field. In Eq. (1), and in the rest of the article, time is normalized to the inverse electron plasma frequency $\omega_{pe}^{-1}$, space is normalized to the Debye length
$\lambda_D$, and velocity is normalized to the electron thermal speed $v_{Te}=\lambda_D\omega_{pe}$. Ions are taken to be motionless, and their only role is to provide a uniform, neutralizing background. Furthermore, periodic boundary conditions
are assumed in $x$. 

As it was mentioned, the present solution of Vlasov equation is based on following the phase points trajectories along which phase-space DF is constant. In order to obtain the phase point trajectory one has to solve the characteristics of the Vlasov equation,
\begin{eqnarray}
&& dx_p/dt = v_p,  \nonumber\\ 
&& dv_p/dt = -E_p, \label{2}
\end{eqnarray}
where subscript ``$p$" stands for ``phase point".

Let us first begin with the free streaming part (the advection term) of the Vlasov equation, $\partial_t f + v \partial_x f = 0$, through the following example. The solution of the advection part at a time $t$ is given as a function of the
initial condition by the relation $f(x,v,t) = f(x-v t,v,0)$. If we consider an initial Maxwellian distribution perturbed by a small perturbation, $f(x,v,0) = 1/\sqrt{2\pi} \exp(-v^2/2)\left[1+ \epsilon \cos(kx)\right]$, then the charge density ($\int_{-\infty}^{+\infty} f dv - 1$) will be given by $\rho(x,t)= \exp(-k^2t^2)\epsilon \cos(kx)$ \cite{Shukri}. The analytical solution is decaying exponentially in time. For the numerical solution, the charge density is calculated at every spatial grid point by summation over all grid points in velocity space. On the Eulerian grid due to equal spacing along the velocity axis $\Delta v$, the initial condition can be reconstructed at recurrence time, $T_R = 2\pi/(k \Delta v)$ \cite{Shukri}.  

Now, let us examine the above analytical solution by the method of characteristics. For the free streaming, we just need to solve $dx/dt = v$. Since in free streaming the velocity of the phase points is constant, the characteristic equation can be exactly solved, that is $x_p^{n+1} = x_p^{n} + v_p t^n$, where the superscript denotes $t = n \Delta t$. We first consider a fixed grid with regular phase points arrangement (Fig. 1a). Second, according to each $x_p$ and $v_p$, its $f_p$ is allocated. Next, $x_p$ is advanced one time step while $v_p$ remains constant. Then, interpolation is performed between the phase points and the fixed grid in phase space by BIS \cite{Kazeminezhad, Bilinear} to obtain $f_g$. Finally, the charge density is obtained by summation (here, Trapezoidal rule) over all grid points in velocity space.        

\begin{figure}
\hskip 3.5cm
\includegraphics[height=4cm,width=4.6cm]{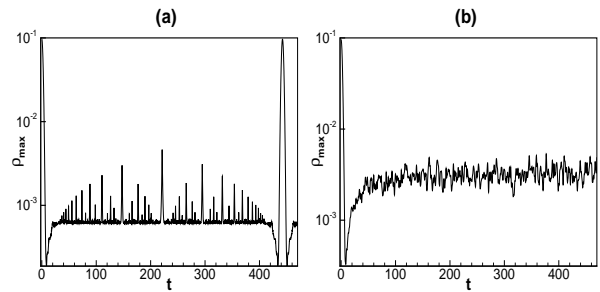} \\
\caption{\label{fig2}The simulation result of the free streaming part of the Vlasov equation with BIS. (a) Regular arrangement (b) Random arrangement along the velocity direction.}
\end{figure}

To compare the simulation result, we choose the parameters similar to Ref. \cite{Shukri}. That is, $\epsilon = 0.1$, with grid points $N_x \times N_v = 16 \times 32$. The length in space is $L = 4\pi$ and in velocity space we use $-5\leq v \leq 5$. The recurrence time of the Eulerian codes, for this case, is $T_R = 38.95$. We put hundred phase points ($10 \times 10$) in each grid cell. The first result of the model was surprising. The recurrence took place at $442.3362$ instead of $38.95$ (Fig. 2a). We realized that contrary to the Eulerian codes, it is not the velocity grid that specifies the recurrence time. 

Thus, the main question is ``How should the recurrence time be calculated?" To answer this question we have to note that it is the evolution of the phase points arrangement that changes the interpolation weighting and therefore $f_g$ (also $\rho$). Thus, as time goes on, the recurrence will take place if there is a possibility to reconstruct the initial phase points arrangement (arrangement at $t=0$). According to Fig. 1a, the velocity spacing in regular arrangement is $dpv$. That means, the smallest velocity for the moving phase point is $dpv$ (in the positive direction) and the other phase points velocities are integer multiples of $dpv$. Since the boundary condition is periodic, if those phase points that their velocity is $dpv$, move a distance $L$ within a time interval $T_R$, the other phase points will move an integer multiples of $L$ within the same time interval. As a result, all the phase points return to their initial positions and the recurrence occurs. Accordingly, in our model, the recurrence time is $T_R = L/dpv$. In another word, there is a much smaller scale $dpv$, in comparison to the grid spacing, along the velocity axis that becomes an important factor in the dynamics. The influence of $dpv$ on the recurrence time is an evidence regarding our claim that fine structures might be more conveniently handled. For the above example, $dpv=0.0284$ that leads to correct recurrence time (Fig. 2a). It is obvious from the Fig. 2a that there are several other recurrences with smaller amplitude. According to our analysis, these smaller amplitude recurrences are due to sub-arrangements  of the phase points, that is, a small group of phase points (not all) has return to their original arrangement.    

\begin{figure}
\hskip 3.5cm
\includegraphics[height=4cm,width=4.6cm]{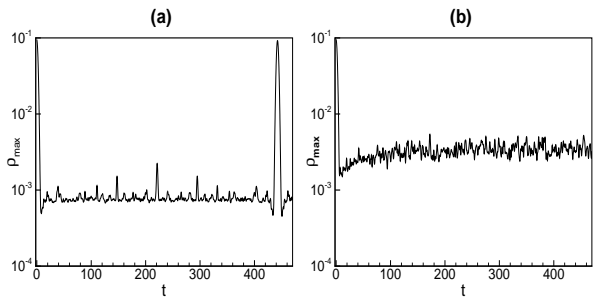} \\
\caption{\label{fig3}The simulation result of the free streaming part of the Vlasov equation with AIS. (a) Regular arrangement (b) Random arrangement along the velocity direction.}
\end{figure}                    
%\vskip -2cm
The latter analysis of the recurrence time is based on two facts. First, the boundary condition is periodic. Second, the velocity of each phase point is an integer multiples of the others. Therefore, by randomizing the phase points velocities (Fig. 1b), we can prevent the occurrence of recurrence (Fig. 2b). To do that, we used a random generator to modify the velocity of each phase point in the range $[-dpv/2,dpv/2]$ (Fig. 1b). Figure 2 depicts the result after randomizing the velocities of the phase points. The recurrence that is supposed to take place at $442.3362$ does not happen. That means our suggestion for the recurrence mechanism works. 

As it was mentioned, increasing the number of phase points necessitates an effective IS that reduces the number of operations while keeping the accuracy ({\it e.g.} in comparison to BIS). For this purpose, we develop a new IS that is almost as accurate as BIS with the difference that it does not use the weighting mechanism.    

Let us call the new IS as average interpolation scheme (AIS). In AIS like BIS, all the phase points should be swept one by one to find those that are located in four cells around a specific grid point (with the important difference that in AIS, finding the host cell of the phase point is the only step, please see below). Now, we denote the distance between the $i$th phase point (in those four cells) and the grid point along $x$ axis by $\Delta x_i$ and along $v$ axis by $\Delta v_i$. Then, by using the Taylor expansion of $f_{pi}(x_g+\Delta x_i,v_g+\Delta v_i)$ around ($x_g,v_g$) and summing over all $J$ phase points, located in four cell, we obtain
\begin{eqnarray}
&&f_g(x_g,v_g) = \frac{1}{J}\sum_{i=1}^J f_{pi}(x_g+\Delta x_i,v_g+\Delta v_i)  \nonumber\\
&&-(\partial_x f_g)\frac{1}{J}\sum_{i=1}^J\Delta x_i - (\partial_v f_g)\frac{1}{J}\sum_{i=1}^J \Delta v_i + O(\Delta^2).\nonumber
\end{eqnarray}     
For an optimum total number of phase points, their initial density in phase space is almost uniform and therefore, $\sum_{i=1}^J\Delta x_i\cong0$ and $\sum_{i=1}^J\Delta  v_i\cong0$. Moreover, the factor of $1/J$ makes the approximation better. According to Liouville's theorem the density of system points in the vicinity of a given system point traveling through phase-space is constant with time. Therefore, the uniformity of phase points density is almost a constant of motion (to the extent of the truncation error). That means,  
\begin{equation}
f_g(x_g,v_g) = \frac{1}{J}\sum_{i=1}^J f_{pi}(x_{pi},v_{pi}) - O(\Delta^2). \nonumber
\end{equation}     
Note, the phase points positions do not explicitly interfere in AIS (as through weighting comes to play in BIS). This is an essential feature of AIS that first makes AIS easily extendable to higher dimensions and second reduces the number of operations. 
\begin{figure}
\hskip -4cm
\includegraphics[height=4cm,width=4.6cm]{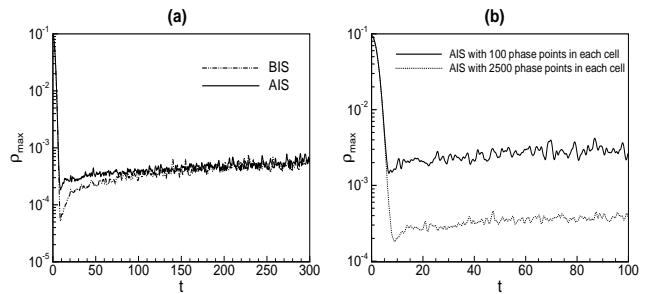} \\
\caption{\label{fig4} (a) Comparison of BIS and AIS. (b) Improvement of the accuracy by increasing the number of phase points.}
\end{figure}                    

Figure 3 demonstrates the simulation of free streaming that is performed using AIS. Fig. 3a shows the result when the initial phase points arrangement is regular. At it is expected a recurrence happen at $t=442.3362$. However, after randomizing the initial phase point arrangements, the recurrence does not take place. 

Now, we first compare the accuracy of BIS and AIS and then show that the accuracy of the method will improve by increasing the number of phase points. In order to compare BIS and AIS, we perform two simulations with these ISs. Both of simulations are initially fed by random phase points arrangement with $2500$ phase points in each cell. Fig. 4(a) exhibits the result. Although, BIS is slightly more accurate (maximum $1\times 10^{-4}$) but AIS interpolates much faster than BIS. It is clear that after initial stage the accuracy of two interpolations becomes almost similar. Note that in this letter we use the simplest scheme with the parameters that are not necessarily the most appropriate ones ($\Delta x = 0.83$ and $\Delta v = 0.3125$. Recall that the accuracy of the interpolation schemes is $O(\Delta^2)$. Therefore, we have not invested on the accuracy and the parameters have chosen to make our results comparable with the results of Ref. \cite{Shukri}. Next, in order to show the improvement of accuracy of the method, we redo the simulation of free-streaming by AIS for two different cases when $100$ and $2500$ phase points are put in each cell. Figure 4(b) demonstrates the results. It is obvious that when we put $2500$ phase points in each cell, the result is one order of magnitude more accurate than the case when the DF is sampled by $100$ phase points within a cell.       

\begin{figure}
%\hskip -1.5cm
\includegraphics[height=4cm,width=8cm]{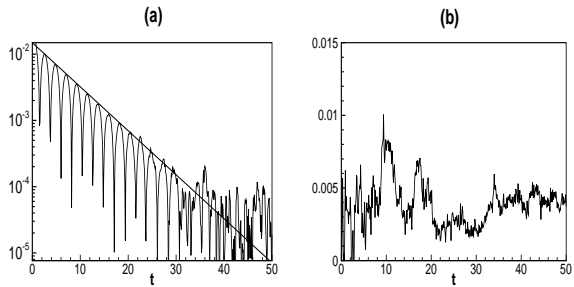} \\
\caption{\label{fig5}The linear Landau damping (a) The exponential decay of the amplitude of the electric field with the damping rate of $\gamma=0.153$ (b) The relative error in the total energy in percent.}
\end{figure}

We now examine classical numerical test of the linear Landau damping. In this case, we start with $f(x,v,0) = 1/\sqrt{2\pi} \exp(-v^2/2)\left[1+ \epsilon \cos(kx)\right]$. The initial arrangement of the phase points is random along $v$-axis and regular along $x$-axis. By AIS, $f_g$ is calculated. Integrating $f_g$ over the velocity space and solving Poisson's equation leads to the electric field on the grid, $E_g$. Poisson's equation is solved by the Fast Fourier Transform. Then, by a Lagrange polynomial IS \cite{Numerical recipes}, $E_g$ is interpolated to the position of phase points to obtain $E_p$. Having $E_p$, Eqs. (2) are solved by the Leapfrog-Trapezoidal scheme. The parameters are $\epsilon = 0.01$, with grid points $N_x \times N_v = 256 \times 256$ and in each cell we put $4$ phase points. The periodic length is $L = 4\pi$, $k=0.5$, $-5\leq v \leq 5$ , and $dt=0.1$. In Fig. 5a, the basic mode of the electric field is plotted against time. It shows the exponential decay of the amplitude of the electric field according to Landau's theory. The damping rate, the slope of straight line, obtained by this method is $\gamma=0.153$ which agrees very well with values predicted by the theory \cite{Bertrand}. Figure 5b exhibits the relative error in the total energy in percent, i.e. (total energy$^{n}$-total energy$^0$)/total energy$^0$ $\times 100$, recalling that the superscript ''$n$" denotes the quantities at  $t=n \Delta t$). As it is seen, the scheme is capable of keeping the energy conservation. 

In summary, we improved the method of characteristics \cite{Kazeminezhad} by increasing the number of phase points in phase space. Naturally, due to phase points dynamics the  development of steep gradients can be managed without enhancing the velocity resolution. 
In order to reduce the number of operation AIS is introduced that is easily extendable to higher dimension.  By AIS and randomizing the phase point arrangement, we could prevent the occurrence of the recurrence effect. The scheme is very similar to PIC simulation with the advantages that it is noise-free with a simpler IS (i.e. weighting is omitted) and that through DF it is much easier to calculate the thermodynamic quantities. Using today's supercomputers, this method appears to be a good alternative to the PIC methods for dealing with strongly nonlinear problems in phase space when little noise and good precision is needed.

We thank F. Kazeminezhad for his continual support and comments, B. Eliasson for his remarks 
on the recurrence effect, S. Kuhn for his helpful discussions, M. Shoucri, I. Hofmann, and E. Sonnendrucker for reviewing the manuscript and their valuable comments. 
%\newpage %Just because of unusual number of tables stacked at end

\end{document}